\documentclass[aps,prb,twocolumn,superscriptaddress,amsmath,amssymb,nolongbibliography]{revtex4-2}
\usepackage{graphicx}
\usepackage{hyperref}
\usepackage{braket}
\usepackage{amsfonts}
\usepackage{physics}
\usepackage{xcolor}
\usepackage[margin=1in]{geometry}
\usepackage{graphicx}

\begin{document}
	
	\title{Pair scattering from time-modulated impurity in the Bose-Hubbard model}
	
	\author{Neda Ahmadi}
	\affiliation{Physikalisches Institut, University of Bonn, Nussallee 12, 53115 Bonn, Germany
	}
	
	\author{Ameneh Sheikhan}
	\affiliation{Physikalisches Institut, University of Bonn, Nussallee 12, 53115 Bonn, Germany
	}
	
	\author{Corinna Kollath}
	\affiliation{Physikalisches Institut, University of Bonn, Nussallee 12, 53115 Bonn, Germany
	}
	
	\date{\today}
	
	\begin{abstract}
	We investigate scattering phenomena in a one-dimensional attractive Bose-Hubbard model with a time-periodically modulated impurity. We analyze both single-particle and pair (doublon) transmission, exploring a range of interaction strengths and drive amplitudes. Our exact numerical results reveal excellent quantitative agreement with analytical predictions in the high-frequency limit. At intermediate and weak attractive interactions, we observe significant pair dissociation and the emergence of dynamically localized single-particle modes. These features are reminiscent of Floquet Bound States in the Continuum (BICs). These findings provide new avenues for engineering controllable quantum transport and localized states in ultracold atom experiments.

	\end{abstract}
	
	\maketitle

	\section{Introduction} 
In the past decade, the application of time-periodic modulation has proven to be an important tool to control many body systems. 
Early on, effects as the dynamical localization of particles in a lattice by the application of a driving potential have been theoretically predicted and investigated\cite{GrossmannHaenggi1991,DunlapKenkre1986,Holthaus1992} and experimentally realized in ultracold atomic gases \cite{LignierArimondo2007}. Following an early proposal \cite{JakschZoller2003}, the periodic driving has  been employed using a two Raman process in order to engineer artificial magnetic fields for ultracold atomic gases~\cite{LinSpielman2009,AidelsburgerBloch2011,MiyakeKetterle2013,AidelsburgerBloch2013,JotzuEsslinger2014,GoldmanDalibard2014,AidelsburgerGoldman2018}.
The driven phase transitions between a superfluid and a Mott-insulator or to superconductors has been investigated~\cite{EckardtHolthaus2005,ZenesiniArimondo2009,PolettiKollath2011,KitamuraAoki2016,SheikhanKollath2020} and controllable tunneling of pairs~\cite{KudoMonteiro2009,MeinertNaegerl2016, KlemmerBergschneider2024} or formation of pair bound states in the continuum ~\cite{DellavalleLonghi2014}.

Theoretically these systems are often described using the Floquet theory~\cite{Shirley1965,BukovPolkovnikov2015,EckardtAnisimovas2015,Eckardt2017}, which treats the periodically modulated quantum systems in terms of quasi-energies, analogous to quasi-momenta in spatially periodic systems. Using Floquet theory facilitates often the understanding of key effects, such as resonance effects or the realization of effective Hamiltonians.

One recent research direction is to engineer the transport properties of strongly correlated particles by using local drives. Such Floquet-engineered impurity sites are used as building blocks for new quantum devices that can be easily adjusted. For example, periodically driven quantum dots and point contacts have been proposed as energy filters~\cite{KohlerHaenggi2005,ThubergEggert2016,ReyesEggert2017,GamayunLychkovskiy2021}, spin filters~\cite{KimSatanin1998,Kim1999,ThubergReyes2017}, or as generators of bound states in the continuum~\cite{LonghiDellavalle2013,AgarwalaSen2017}.

Additionally to these single particle effects, the interplay between interactions and time-periodic localized driving fields can lead to very complex phenomena and has been known as a powerful tool for engineering novel quantum phenomena such as single or pair filters \cite{HuebnerSheikhan2022,HuebnerSheikhan2023}. The pair filters or Floquet-Hubbard bound states rely on the formation and propagation of tightly bound on-site pairs (doublons), which become stable by strong attractive interactions. These pairs then show very distinct transport properties from the single particles in the presence of the driven impurity. The previous theoretical predictions  \cite{HuebnerSheikhan2022,HuebnerSheikhan2023} were based on perturbative expansions as the (Floquet-) Schrieffer-Wolff transformation. These treatments assume that the pairs are stable and derive an effective pair Hamiltonian.  Thus, an interesting question which needs to be addressed going beyond the perturbative approach is whether such pairs remain bound or dissociate in the presence of the driven impurity. 

In this paper we address this question using exact calculations of the pair dynamics. We are interested in the real-time dynamics of a pair wave packet injected into a one-dimensional Bose-Hubbard chain in the presence of a time-periodically modulated impurity potential at the center of the chain. By combining numerically exact simulations with effective analytical descriptions in the strongly interacting regime, we investigate the dynamics of the bound pair. We study whether the pair is reflected or transmitted and whether the pair remains intact, undergoes partial dissociation, or completely breaks into single particles. One focus is how the modulation amplitude and frequency affects the transport of both single particles and  pairs.

The paper is organized as follows. In Section~\ref{sec:ModelMethod}, we present the model Hamiltonian and outline the wave-packet formalism used to probe transport properties. Section~\ref{sec:SingleParticle} focuses on the dynamics of single-particle excitations, validating our numerical approach by comparing with analytical results from effective Floquet theory. Section~\ref{sec:DoublonTransport} explores the transport of on-site pairs in various driving regimes, highlighting phenomena such as pair transport suppression, pair breaking, and Floquet-induced localization. We provide a detailed analysis of real-space density and correlation profiles to understand this effect better. Finally, in Section~\ref{sec:Conclusion}, we summarize our main findings and discuss potential extensions.
	
	\section{Model and Methods}\label{sec:ModelMethod}
	We consider a one-dimensional Bose-Hubbard model with a time-dependent impurity that is driven periodically. The impurity is located at site $j=0$. We describe the system by the Hamiltonian:
	\begin{eqnarray}\label{Eq:Hamiltonian}
	\nonumber	&H(t)& = -J \sum_{j=-\L/2+1}^{L/2-1} \left( a^\dagger_j a^{\phantom{\dagger}}_{j+1} + \text{h.c.} \right) \\
	&+& \frac{U}{2} \sum_{j=-L/2+1}^{L/2} n_j (n_j - 1) 
	+ \lambda \hbar\omega \cos(\omega t) n_0.
	\end{eqnarray} 
	Here, $a^{\phantom{\dagger}}_j(a^{\dagger}_j)$  is the bosonic annihilation(creation) operator at site $j$, $n_j =a^{\dagger}_ja^{\phantom{\dagger}}_j$ is the density operator and $L$ is the number of sites.
Here $J$ denotes the hopping amplitude and is taken as the energy unit, $U<0$ is the attractive onsite interaction between the bosonic particles. For the ground state, the attractive interaction would lead to an unstable state for many bosons. However, here we consider a single particle or a single pair of bosons for which the attractive situation is stable. The impurity is driven periodically in time with the driving frequency of $\omega$ and the dimensionless driving amplitude $\lambda$. The lattice spacing is $d$.

Since we are interested in the propagation of wave packets through the impurity, we initialize the system by creating a wave packet on top of the vacuum state,

\( \left| 0 \right\rangle \). The vacuum state represents an empty lattice with no particles.
A Gaussian wave packet is created at a site \( j_0 < 0 \), well to the left of the impurity. This wave packet is constructed with a mean momentum \( k_0 \), where $k_0>0$ and spatial width \( \sigma \), with the action of the operator
\begin{equation}
	W_{\alpha}^{\dagger}(k_0) = A \sum_j e^{-\frac{(j - j_0)^2 d^2}{2\sigma^2}} e^{i k_0 d j} \mathcal{O}_{\alpha, j}^{\dagger}
\end{equation}
on the vacuum state. Here, \( \alpha \in \{s, p\} \) denotes the nature of the excitation, with \( \mathcal{O}_{s,j}^{\dagger} = a_j^{\dagger} \) for a single-particle excitation and \( \mathcal{O}_{p,j}^{\dagger} = a_j^{\dagger} a_j^{\dagger} \) for an onsite pair (doublon) excitation. The constant \( A \) ensures proper normalization of the wave packets.

The initial state of the system with the created wave packet is given by
\begin{equation}
	\left| \psi_{\alpha}(t = 0) \right\rangle = W_{\alpha}^{\dagger}(k_0) \left| 0 \right\rangle.
\end{equation}

To distinguish between paired and unpaired particles in our system, we compute two observables: the total local density $\langle n_j \rangle$ and the local two-particle density $\langle n_j(n_j - 1) \rangle$, which we refer to as the pair density. Since the system contains at most two bosons, the local occupation number $n_j$ can only take the values 0, 1, or 2. This makes $\langle n_j(n_j - 1) \rangle$ a good indicator of on-site pairing, as it vanishes when $n_j = 0$ or $1$, and becomes non-zero only when $n_j = 2$. The total or pair transmission coefficient, which quantifies the fraction of the wave packet transmitted across the impurity region, is computed at sufficiently long  times \( t_{\text{final}} \) as:
\begin{eqnarray} \label{Eq:transport}
	\nonumber T^{n}_{\alpha}(k_0) &=&\frac{1}{N_\alpha} \sum_{j > 0} \left\langle \psi_\alpha(t_{\text{final}}) \middle| n_j \middle| \psi_\alpha(t_{\text{final}}) \right\rangle,\\
\nonumber	T^{n(n-1)}_{p}(k_0) &=&\frac{1}{N_p}  \sum_{j > 0} \left\langle \psi_p(t_{\text{final}}) \middle| n_j(n_j-1) \middle| \psi_p(t_{\text{final}}) \right\rangle.\\
\end{eqnarray}
where the sum over \( j > 0 \) covers the sites to the right of the impurity, $N_s=1$ and $N_p=2$. The pair transmission is non-zero only if the initial wavepacket contains a pair, thus we used $\alpha=p$.

\section{Results}	

\begin{figure}
\centering
\includegraphics[width=\linewidth]{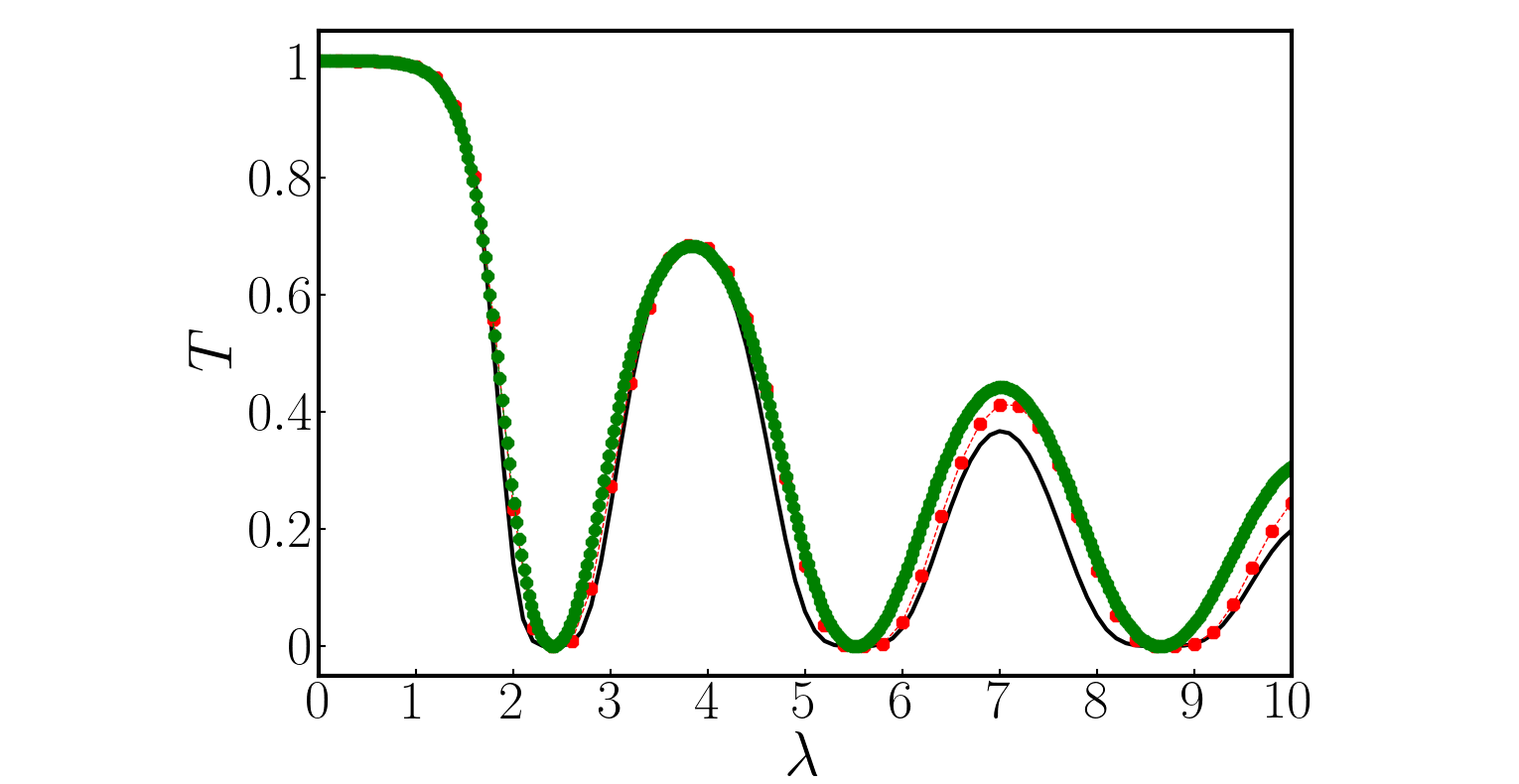}
\caption{Single particle transmission as a function of the driving amplitude $\lambda$. The numerical results $T_s^n$ (red dots) are compared with the analytic results $T_{s,\hbar\omega\gg J}^n(k_0)$  in the high frequency limit Eq.~\ref{Eq:singleTransmission}
 (solid black line) and the averaged analytic results Eq.~\ref{Eq:integralOverT} (green dots). In the simulation we use $L=128$, $\hbar \omega = 50J$, $k_0 d=1.7$, $j_0=-30$ and $\sigma=12d$.} 
\label{fig:transportsingle}
\end{figure}

\subsection{Single Particle Transmission}\label{sec:SingleParticle}
In this section, we focus on the transport properties of a single-particle excitation propagating through the periodically driven impurity. This situation has been investigated previously in several publications \cite{HuebnerSheikhan2022,HuebnerSheikhan2023,ReyesEggert2017}. In particular, analytical predictions are found in the high frequency limit. We summarize here for completeness some of these previous findings to contrast it to the pair transmission later and use it as a check for our numerical results. 

In the high frequency regime which has been studied in Ref. \cite{ReyesEggert2017,HuebnerSheikhan2022,HuebnerSheikhan2023}
one can use the high frequency expansion \cite{EckardtAnisimovas2015}, and an effective time independent Floquet Hamiltonian is derived given by 
\begin{eqnarray}\label{Eq:HamiltonianS}
	\nonumber H_{\text {eff}}^s &=& -J\sum_{j\neq -1, 0}\left(a_j^{\dagger}a_{j+1} + h.c. \right) \\
	&-&J\mathcal{J}_0\left( \lambda \right) \sum_{j= -1, 0}\left(a_j^{\dagger}a_{j+1} + h.c. \right).
\end{eqnarray}

Here $\mathcal{J}_0$ is the Bessel function of the first kind. This effective Hamiltonian shows that the driving effectively results in a reduced hopping to and from the impurity site. At the zeros of the Bessel function, this hopping is even completely suppressed and single particles cannot traverse the impurity. As shown in Ref. \cite{ReyesEggert2017}, the transmission of an incoming single particle with momentum $k_0$ through the impurity is:
\begin{equation}\label{Eq:singleTransmission}
	T_{s,\hbar\omega\gg J}^n(k_0) = \frac{1}{1+(\frac{1}{\mathcal{J}_0(\lambda)^2}-1)^2\cot^2 k_0 d}.
\end{equation}
We see that the transmission of the single particle depends on the momentum and on the rescaled driving amplitude $\lambda$. To better compare our simulation with the analytical theory, we compute an averaged analytical transmission by integrating over the momentum distribution of the initial Gaussian wave packet. The transmission for a single momentum component \( k_0 \) is given by the analytical expression \( T_{s,\hbar\omega\gg J}^n(k_0)  \) (Eq.~\ref{Eq:singleTransmission}). The integrated analytical transmission is defined as:
\begin{equation}\label{Eq:integralOverT}
	\bar{T}^s(k_0) = \int_{-\frac{\pi}{d}}^{\frac{\pi}{d}} \frac{dk}{\sqrt{2\pi} \sigma_k} \exp\left[-\frac{(k - k_0)^2}{2\sigma_k^2}\right] 	T_{s,\hbar\omega\gg J}^n(k_0) \, 
\end{equation}
where \( \sigma_k = \frac{1}{\sigma } \) is the momentum-space width of the wave packet. This result has been shown by green dots in Fig.~\ref{fig:transportsingle}

In order to benchmark our simulations, we also compare the analytical results~\cite{ReyesEggert2017} with the numerical results obtained from time-dependent matrix product state (tMPS) simulations~\cite{DaleyVidal2004,WhiteFeiguin2004} implemented using the ITensor library~\cite{itensor}. Due to the low particle number these simulations become exact for a very low bond dimension. Since in the experiments the preparation of a single momentum state is involved, in our simulations, a Gaussian single particle wave packet is initialized on the left side of the chain described by $\psi_s(t=0)$. In the following this wave packet evolves freely under the Hamiltonian Eq.~\ref{Eq:Hamiltonian} and the evolution is calculated modeling a corresponding experiment. The single particle wave packet propagates towards the impurity site. For our simulation we have considered a chain consisting of $L=128$ sites. The time evolution is performed with a time step of $\Delta t=0.01\hbar/J$, which has been verified to give accurate results. The simulations are run up to a final time of $t_{\text{final}}=30\hbar/J $ to ensure the wave packet has fully interacted with the impurity and the transmitted/reflected parts are well-separated.

Once the wave packet has crossed the impurity region, the transmission coefficient is computed using Eq.~\ref{Eq:transport}. In Fig.~\ref{fig:transportsingle} we compare our simulation result with the analytical result (Eq.~\ref{Eq:integralOverT}) obtained for the high frequency regime $(\hbar\omega \gg J)$. As is observed from Fig.~\ref{fig:transportsingle}, there is a good overall agreement between our tMPS simulations (red squares) and the analytical exact results (black stars) for single-particle transmission. In particular, the position of the maxima and minima is well reproduced. The minima show the suppression of single-particle transport at specific $\lambda$, consistent with the coherent destruction of tunneling. This effect emerges when the Bessel function $\mathcal{J}_0(\lambda)$ vanishes, leading to effective suppression of tunneling through the driven impurity site~\cite{ReyesEggert2017,HuebnerSheikhan2022,HuebnerSheikhan2023,EckardtAnisimovas2015,Eckardt2017}.
The remaining very small deviations between the numerical and analytical results we attribute to the choice of the driving frequency, which is still not sufficiently high to make the high frequency approximation exact. We have verified that choosing a lower frequency the deviations become larger. Beyond the suppression points caused by coherent suppression of tunneling, the single-particle transmission exhibits peaks at other values of $\lambda$. At these points, periodic driving restores the effective tunneling amplitude across the impurity site. This behavior is governed by the Bessel-function modulation of the hopping term and is consistent with the structure of the effective Floquet Hamiltonian.

\subsection{Pair Transmission}
\label{sec:DoublonTransport}
In this section we study the transport of a pair of particles created by the attractive onsite interaction. First we revise in Sec.~\ref{sec:SchriefferWolff} previous perturbative results for the small tunneling regime, i.e. $|U|,\hbar \omega \gg J$ on the transmission and present then in section \ref{subsec:pairDMRG} our numerical results which go beyond this limit. 

\subsubsection{Floquet-Schrieffer-Wolff transformation}
\label{sec:SchriefferWolff}
The perturbative results obtained from the Floquet-Schrieffer-Wolff transformation\cite{BukovPolkovnikov2016}  predict for the considered model already a very interesting behaviour of the pair transmission \cite{HuebnerSheikhan2023}. Using the Floquet-Schrieffer Wolff transformation which is valid when $\left| U\right| ,\hbar\omega \gg J$ and the interaction is not commensurate with the driving frequency (i.e. $U\neq l\hbar \omega$ where $l$ is an integer number), one obtains an effective Hamiltonian for attractive pairs given by
\begin{eqnarray}\label{Eq:analyticpairtransmisssion}
	\nonumber H_{\text{eff}}^p &=& -J_p \left[ \sum_{j \neq -1, 0} \left( \eta_j^{+} \eta_{j+1}^{-} + \text{h.c.} \right) \right. \\
	\nonumber &+& \gamma_p \sum_{j =-1, 0} \left( \eta_j^{+} \eta_{j+1}^{-} + \text{h.c.} \right) \\
	&+& \left. \mu_p \left( \eta_{-1}^{+} \eta_{-1}^{-} + 2\eta_0^{+} \eta_{0}^{-} + \eta_{+1}^{+} \eta_{+1}^{-} \right) \right]
\end{eqnarray}
where $\eta_j^{+}= a_j^{\dagger} a_j^{\dagger}$ and $\eta_j^{-}= a_j a_j$ are pair creation and annihilation operators, respectively. The effective pair Hamiltonian shows a tunneling of the pairs with the effective pair hopping amplitude $J_P=\frac{2J^2}{\left| U\right| }$ within the chain which is well known from the standard Schrieffer-Wolff transformation of the attractive Bose-Hubbard model. The tunneling to and from the impurity has a different amplitude which strongly depends on the periodic modulation and is given by  
$$\gamma_p=\sum_{l=-\infty}^{\infty}\frac{\frac{\left| U\right| }{\hbar\omega}(-1)^l\mathcal{J}_l(\lambda)^2}{\frac{\left| U\right|}{ \hbar\omega}-l}.$$
where $\mathcal{J}_l$ is the $l-$th Bessel function. 
Additionally, the pairs feel a triangular potential on the impurity and its neighbouring sites given by
$$\mu_p=\sum_{l=-\infty}^{\infty}\frac{\frac{\left| U\right| }{\hbar\omega}\mathcal{J}_l(\lambda)^2}{\frac{\left| U\right|}{\hbar \omega}-l}-1.$$ 
Within this picture the pair is subjected to a scattering at a region of three sites with scaled tunneling amplitude and a triangular potential.
The momentum-dependent pair transmission  \cite{HuebnerSheikhan2022} through the impurity results in
\begin{eqnarray}\label{Eq:pairexacttrans}
 \nonumber	&&T^{\textrm{FSW}}_p(k_0) = \frac{\gamma_p^4}{1+\left( \frac{\cos(k_0d)-\mu_p}{\sin(k_0d)}\right) ^2}\times\\
 & & \nonumber \frac{1}{\left[(\cos k_0d-\mu_p)^2-\gamma_p^2\right] ^2 +(\cos k_0d-\mu_p)^2\sin^2 k_0d}.\\
\end{eqnarray}
In Fig.~\ref{fig:transportpair} this perturbative result is shown with a black line.
It shows that its behaviour is very different than the single particle behaviour (green line). In particular, there exist values of $\lambda$ for which the single particle transmission almost vanishes, but the pair transmission is relatively high. Previously it had been suggested to use this as a filter for the pairs/single particles, respectively\cite{HuebnerSheikhan2022,HuebnerSheikhan2023}.

\subsubsection{Numerical results on pair transmission}
\label{subsec:pairDMRG}

\begin{figure}[t]
	\centering
	\includegraphics[width=\linewidth]{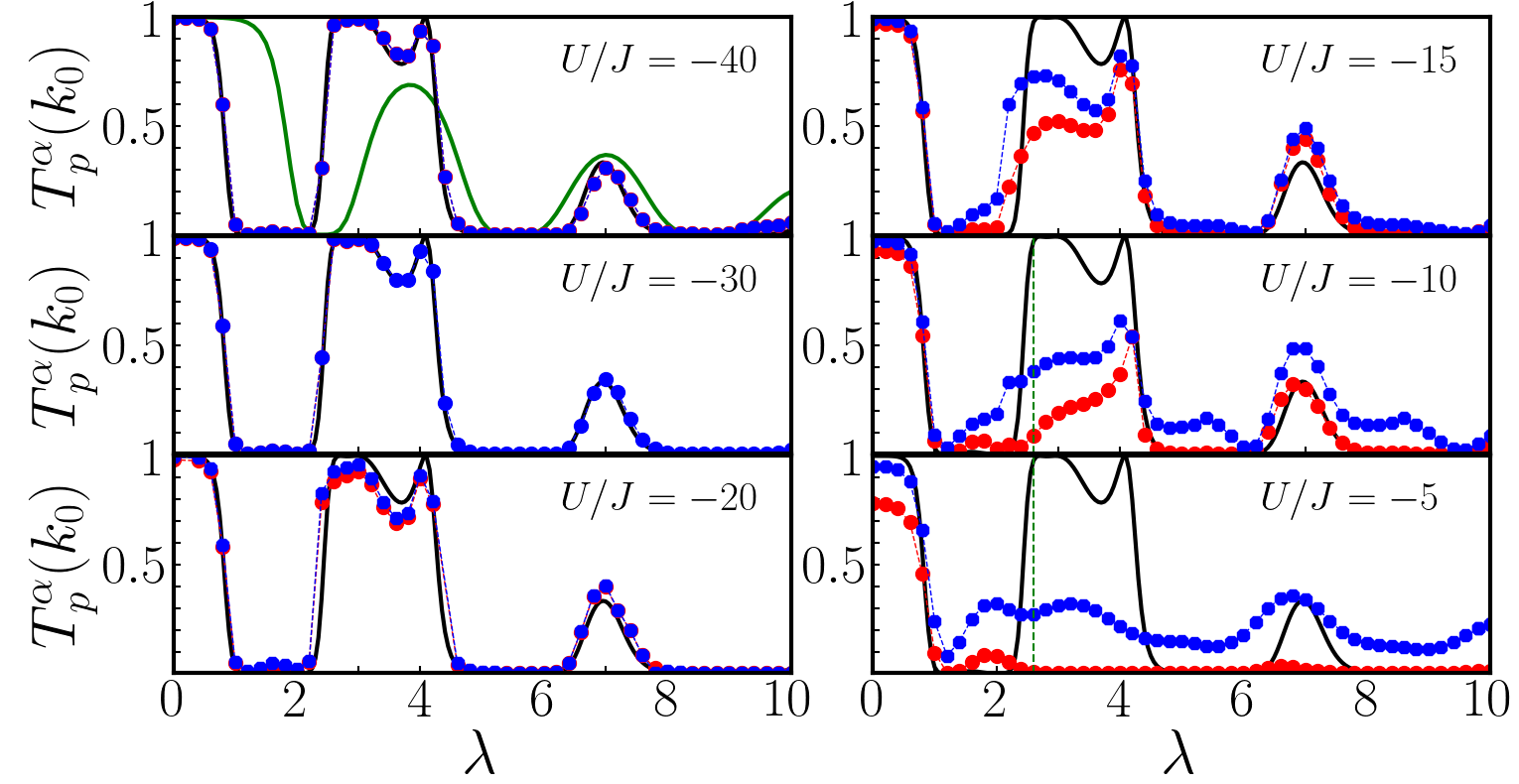}
	\caption{Transmission  $T^{\alpha}_p(k_0)$ as a function of the driving amplitude $\lambda$ for different interaction strengths when an on site pair has been created. Solid black line: results obtained using the Floquet-Schrieffer Wolff transformation ($\alpha=$FSW) given in Eq.~(\ref{Eq:pairexacttrans}), blue circles: numerical result for the total transmission $\alpha=n$ (Eq.~\ref{Eq:transport}), 
          red circles: numerical result for pair transmission $\alpha=n(n-1)$ (Eq.~\ref{Eq:transport}), solid green line: the analytic results of single particle transmission in the high frequency limit (Eq.~\ref{Eq:singleTransmission})  for comparison. 
          In our simulation we used $\left| U\right| /\hbar\omega =2.62 $, $L= 100$, $\sigma=12d$ and $ k_0 d=1.7$, $j_0=-25$, $\Delta t= 0.01\hbar/J$ and the maximum bond dimension is considered is 60. 
	 }
	\label{fig:transportpair}
\end{figure}

\begin{figure}[h]
	\centering
	\includegraphics[width=0.9\linewidth]{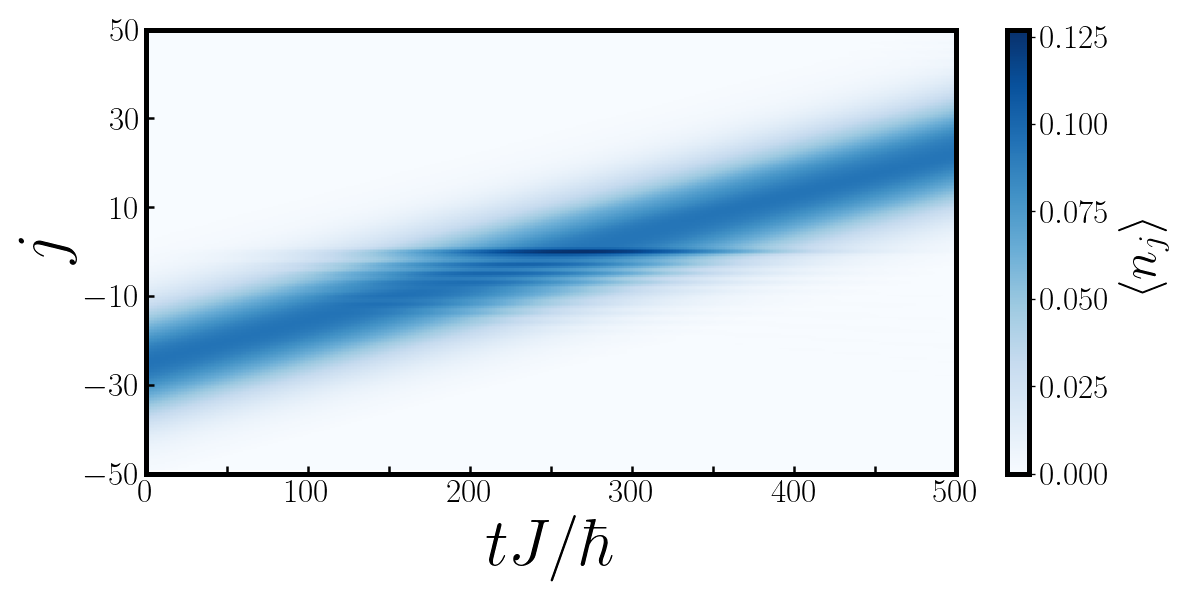}

	\caption{Evolution of the total density $\langle n_j \rangle$ of a pair wave packet as a function of site index $j$ and time $t$ for $U/J = -40$, $\lambda = 0.4$, $\left| U\right| /\hbar\omega =2.62 $ , $L= 100$, $\sigma=12d$, $ k_0 d=1.7$, $j_0=-25$, $\Delta t= 0.01\hbar/J$. The pair propagation is hardly affected by the impurity, and all particles are transmitted.}
	\label{fig:1propagationU40}
\end{figure}

\begin{figure}[h]
	\centering
	\includegraphics[width=0.9\linewidth]{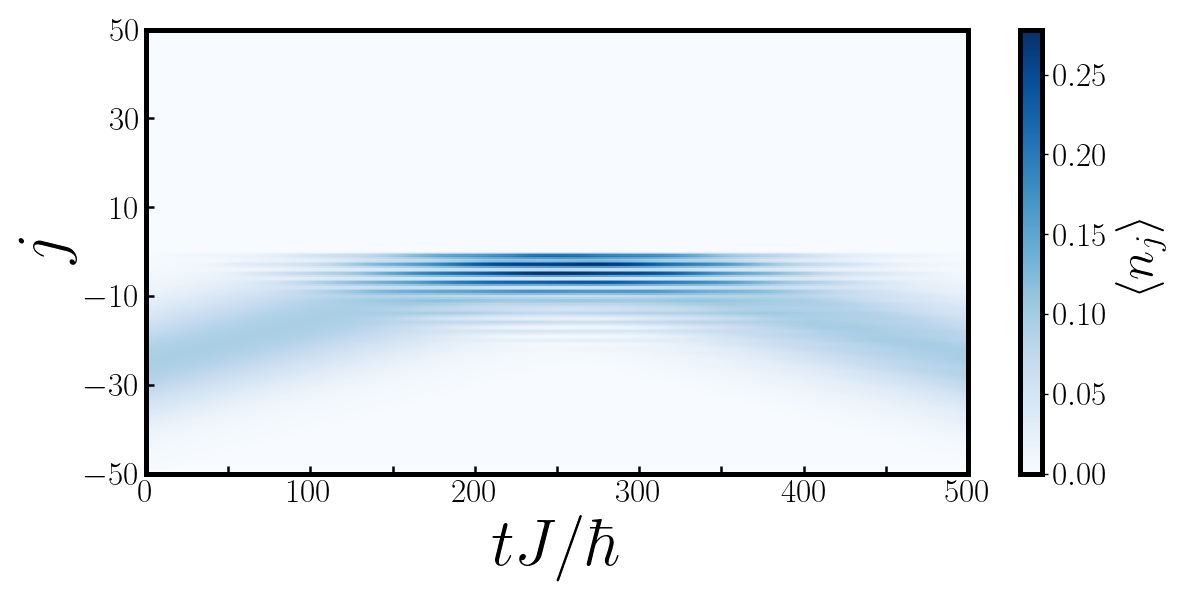}
	\caption{Evolution of the total density $\langle n_j \rangle$ of a pair wave packet as a function of site index $j$ and time $t$ for $U/J = -40$, $\lambda = 6$, $\left| U\right| /\hbar\omega =2.62 $ , $L= 100$, $\sigma=12d$, $ k_0 d=1.7$, $j_0=-25$, $\Delta t= 0.01\hbar/J$. The incoming pair is almost completely reflected at the driven impurity.	}
	\label{fig:2propagationU40}
\end{figure}

Since the results presented in the previous section~\ref{sec:SchriefferWolff} rely on a perturbative treatment of the hopping, in the following we numerically verify these results and then go beyond the regime of validity by lowering the ratio of interaction strength to hopping. 

We create a Gaussian wave-packet of onsite pairs, i.e., $\vert \psi_p(t=0)\rangle$ and let it evolve under the Hamiltonian $H(t)$ given in Eq. \ref{Eq:Hamiltonian}.
Like the single particle wave packet, the pair wave packet moves toward the impurity. When it reaches the impurity, the wave packet is either reflected or transmitted depending on the different parameters such as interaction strength, frequency and drive amplitude. Such an evolution is shown in Fig.~\ref{fig:1propagationU40} for an almost total transmission and in Fig.~\ref{fig:2propagationU40} for the case of an almost complete reflection. At later times most of the wave packet has left the impurity site and we determine the transmission coefficient at such a late time using Eq.~\ref{Eq:transport} for various interaction strengths
fixing the ratio of $U/\hbar\omega=2.62$. The results are shown in Fig.~\ref{fig:transportpair}. The blue circles show the results for the total particle transmission $T^n_p$, while the red circles represent pair (doublon) transmission $T^{n(n-1)}_p$, extracted from the pair density.

 For large interactions $U/J=-40,-30$, the agreement between total particle transmission and pair transmission indicates that the pairs propagate as stable composite particles. No significant pair-breaking is observed for any drive amplitude.
Additionally, the simulation results for these large interactions agree very well with the analytical result by the Floquet-Schrieffer-Wolff transformation Eq. \ref{Eq:pairexacttrans} justifying the perturbative treatment in this parameter regime. As the interaction strength decreases (e.g., \( U/J = -20, -15 \)), the driving frequency \( \hbar\omega \) becomes comparable to the single-particle bandwidth (\( \sim 4J \)), placing the system in an intermediate-frequency regime. For even weaker interactions (e.g., \( U/J = -10, -5 \)), the driving frequency can become even smaller than the single particle bandwidth. We see in Fig.~\ref{fig:transportpair} that the numerical results for these interaction strength deviate strongly from the analytical Floquet-Schrieffer-Wolff prediction (Eq. \ref{Eq:pairexacttrans}). The total particle transmission reduces in particular close to the large maximum between $\lambda =2$ and $\lambda =5$. At the same time, the pair transmission shows an even more drastic decrease, which suggests that in this regime, pair-breaking becomes increasingly likely. The origin for the pair breaking stems from two different processes: the pair breaking in the absence of the impurity and the pair breaking due to the impurity. Whereas in the low driving regime $\lambda \ll 1$ and in the regime of low transmission, the deviations from the Floquet-Schrieffer Wolff results seem relatively small, the largest effects can be seen at intermediate and large driving in the transmission peaks. This suggests that pair breaking takes place upon interacting with the impurity~\cite{HuebnerSheikhan2022,KlemmerBergschneider2024}. We will discuss this process in more detail in section \ref{sec:u=-10pairbreaking}.

\subsection{Pair-Breaking Dynamics and Localized Density Profiles at low interaction strength} 
\label{sec:u=-10pairbreaking}

\begin{figure*}
	\centering
	\includegraphics[width=0.4\linewidth]{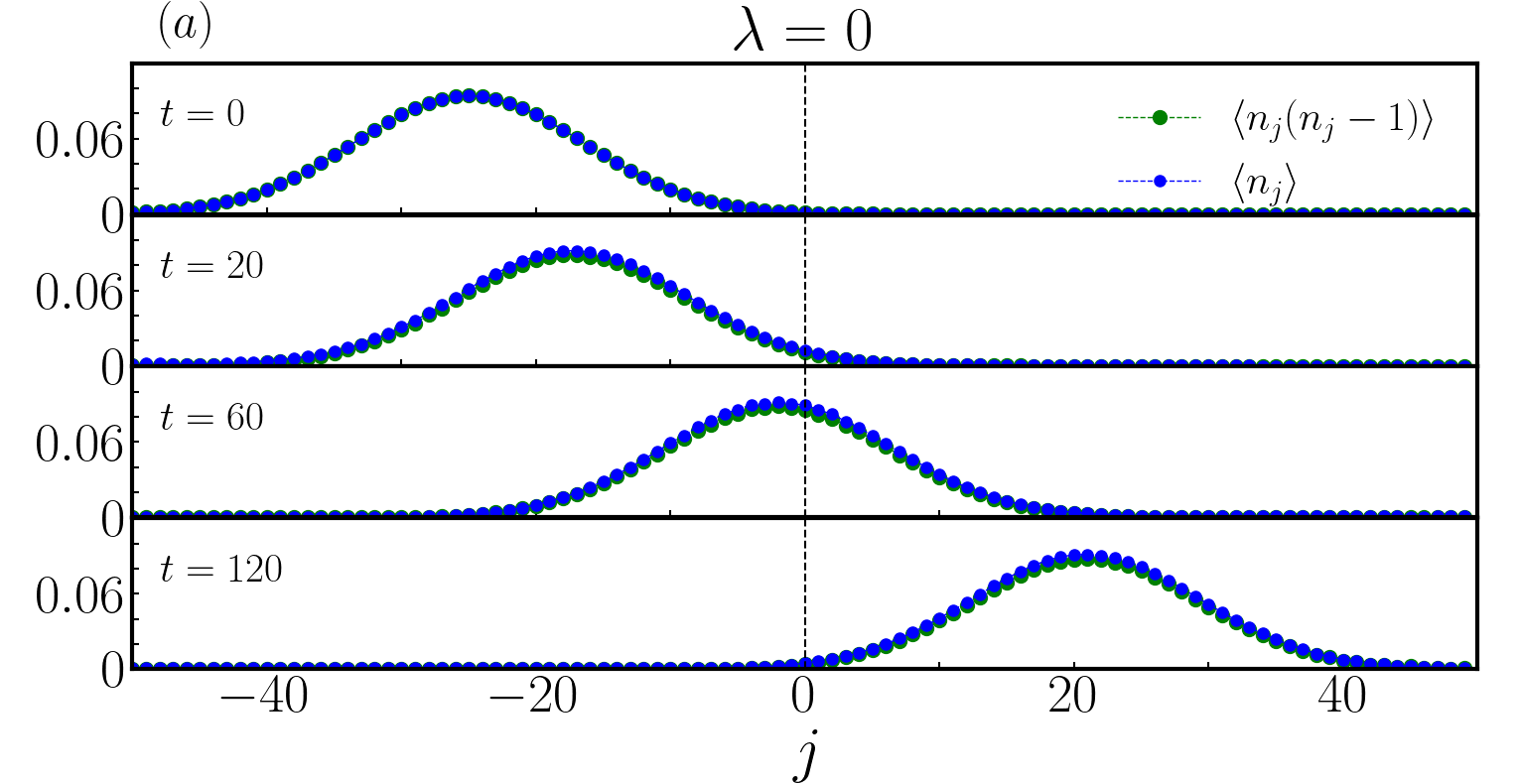}
	\includegraphics[width=0.4\linewidth]{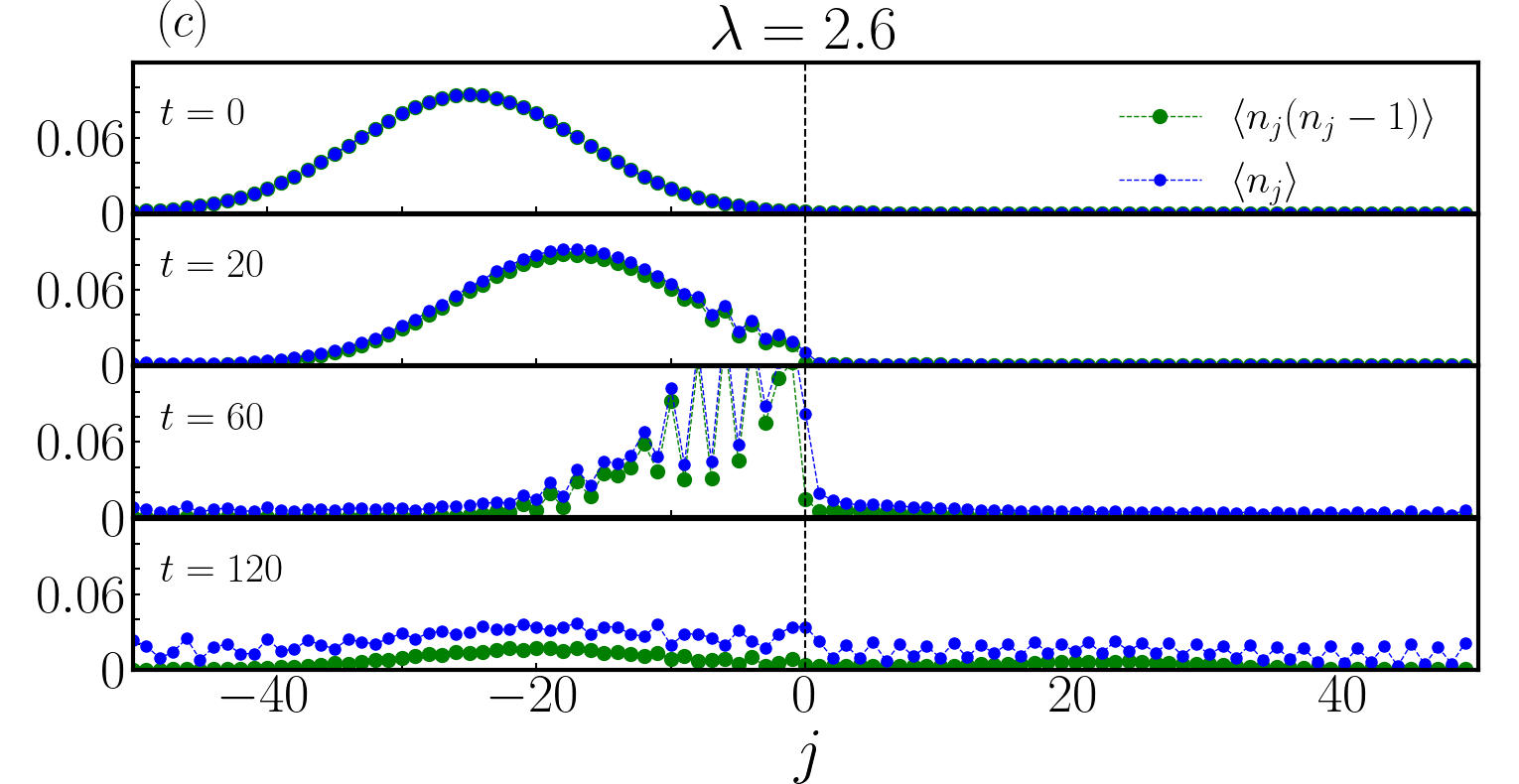}\\
	\includegraphics[width=0.4\linewidth]{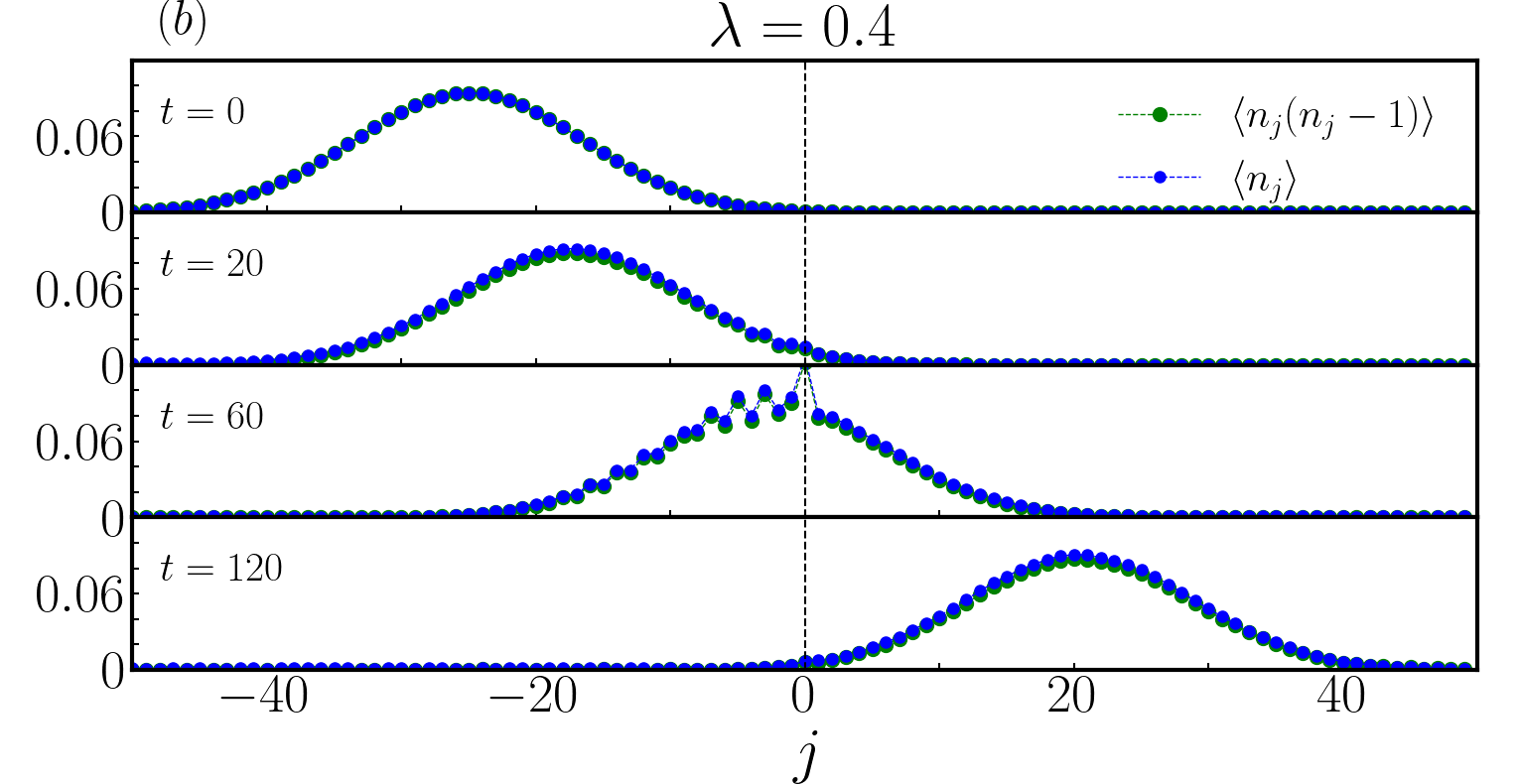}
	\includegraphics[width=0.4\linewidth]{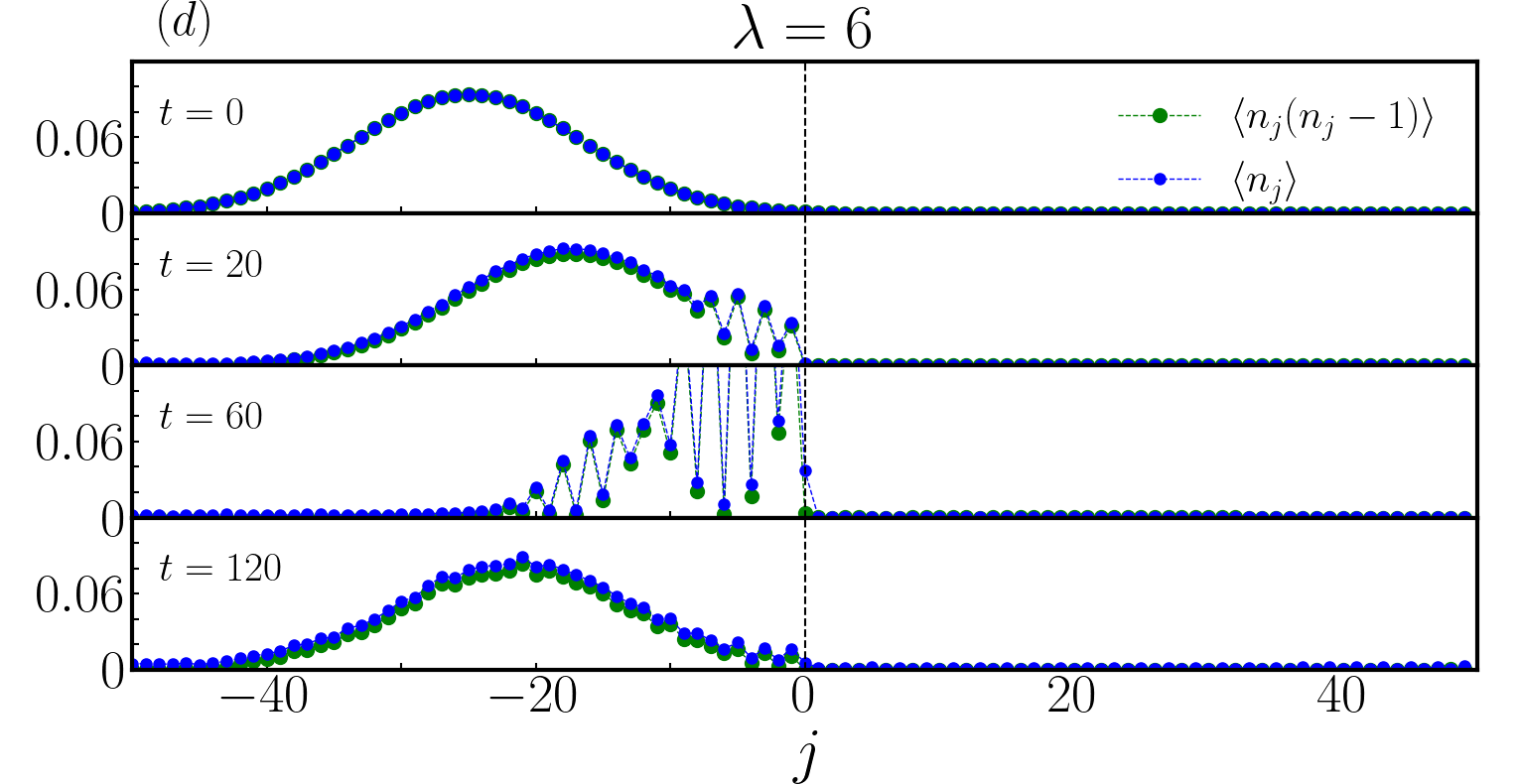}

	\caption{
	Site-resolved single-particle density $\langle n_j \rangle$ (blue) and pair density $\langle n_j(n_j - 1) \rangle$ (green) of propagating pair wave packet at various times for
	(a)  $\lambda = 0.0$, (b)  $\lambda = 0.4$, (c) $\lambda = 2.6$, (d)  $\lambda = 6$. The parameters chosen are $U/J = -10$,  $\left| U\right| /\hbar\omega =2.62 $ ,
	$L= 100$, $\sigma=12d$, $ k_0 d=1.7$, $\Delta t= 0.01\hbar/J$.       
	}
	\label{fig:timecutx}
\end{figure*}
\begin{figure}[h]
	\centering
	\includegraphics[width=0.9\linewidth]{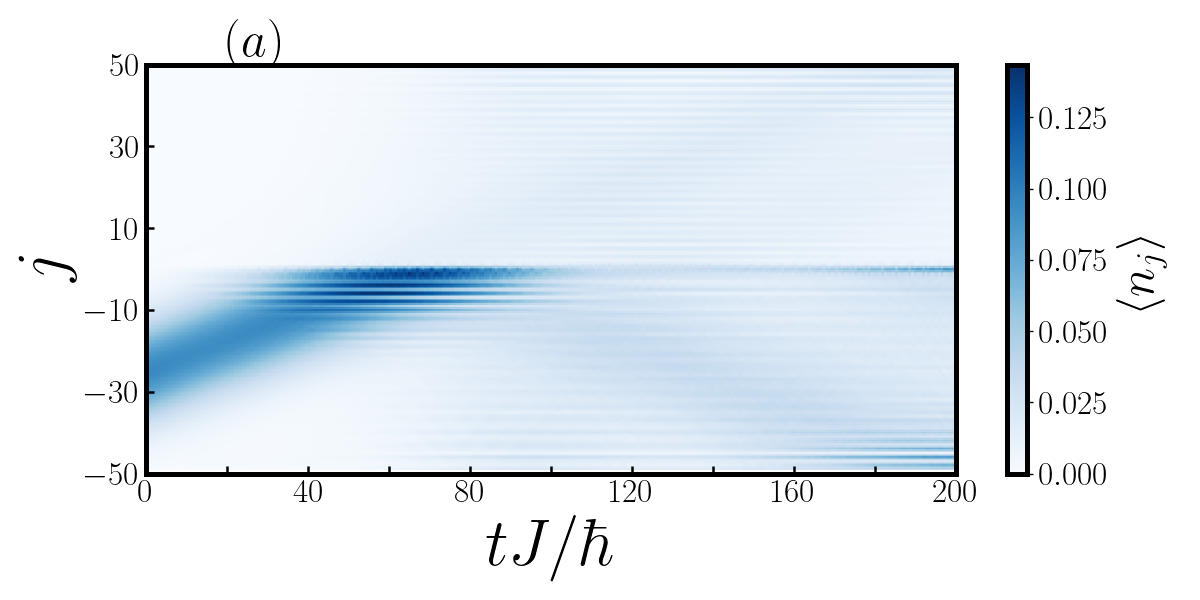}
	\includegraphics[width=0.9\linewidth]{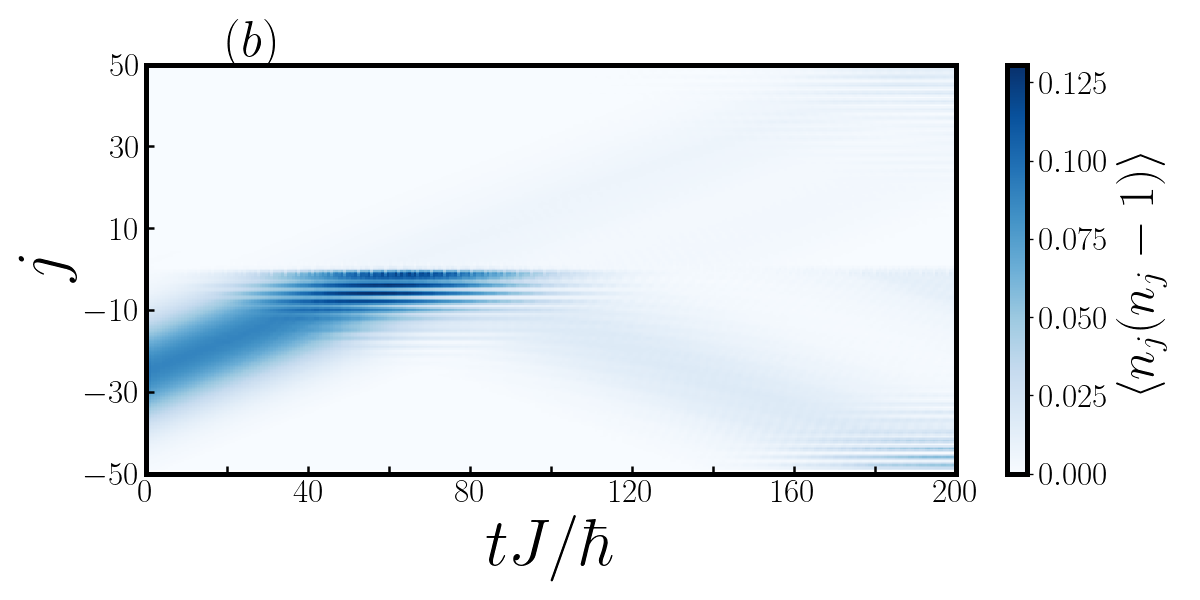}
	\includegraphics[width=0.9\linewidth]{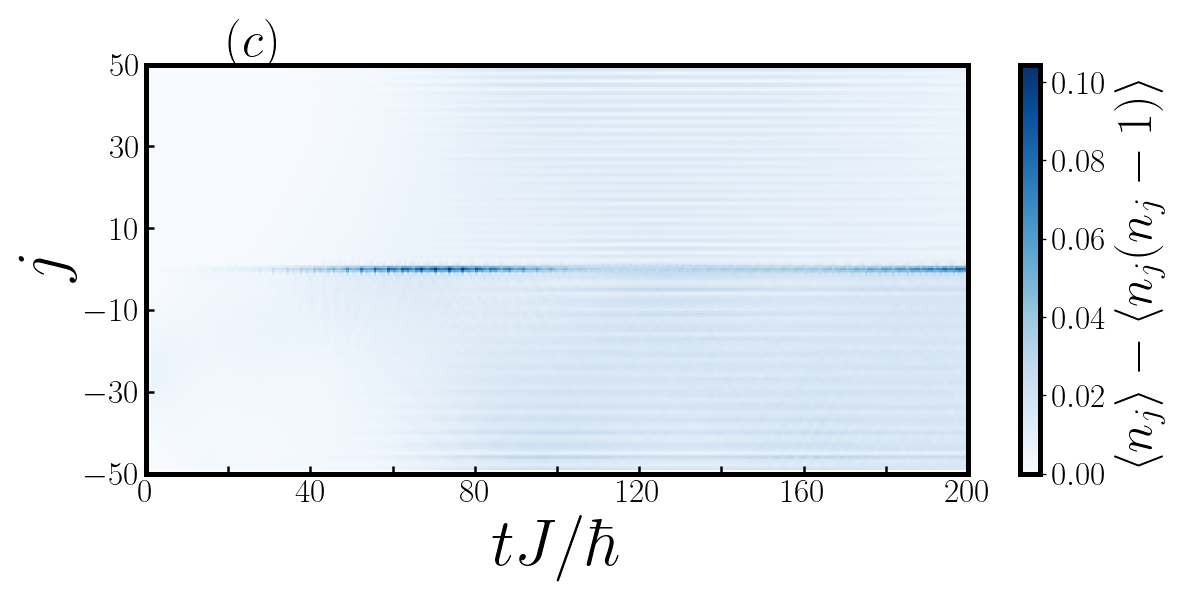}
	\caption{
	Evolution of the (a) total density $\langle n_j \rangle$, (b) the pair density $\langle n_j(n_j - 1) \rangle$, and (c) difference $\langle n_j \rangle - \langle n_j(n_j -1) \rangle$ in time for $U/J = -10$, $\lambda = 2.6$,  $\left| U\right| /\hbar\omega =2.62 $ , $L= 100$, $\sigma=12d$, $ k_0 d=1.7$, $j_0=-25$, $\Delta t= 0.01\hbar/J$. 
	}
	\label{fig:U-10-density}
\end{figure}

To gain further insight into the transport and the origin of the pair breaking, we analyze the site-resolved densities for $U/J = -10$. The plots in Fig.~\ref{fig:timecutx} display the total density $\langle n_j \rangle$ (blue) and the pair density $\langle n_j(n_j - 1) \rangle$ (green) as a function of site index $j$, at some special times  for four driving amplitudes $\lambda = 0.0$, $0.4$, $2.6$, and $6$. We have chosen the specific times when: a) the wave packet has been created b) the wave packet is moving toward the impurity but not reaching it c) the wave packet is exactly at the impurity site d) the wave packet is transmitted or reflected from the impurity. 

In order to see whether the dissociation of the pairs takes also place in the absence of the periodic driving, the impurity potential is effectively switched off  (\(\lambda = 0\)), resulting in a homogeneous lattice. As shown in the top-left panel of Fig.~\ref{fig:timecutx}, both the total and pair densities propagate. In particular, the two curves for the total density and the pair density can hardly be distinguished which means that there is almost no pair breaking taking place. This case serves as a reference for undisturbed pair dynamics, against which the effects of periodic driving can be compared.

As shown in bottom-left panel of Fig.~\ref{fig:timecutx} at $\lambda=0.4$, the wave packet is almost totally transmitted through the impurity site without significant distortion. However, close to the site of the impurity a clear disturbance of the wave packet can be seen, where strong oscillations in the density occur. The total particle and pair densities remain as close as for the non-driven case, indicating that the incoming pair remains bound during propagation. The impurity acts as a weak perturbation in this low-drive regime. This behavior is consistent with expectations from the high frequency prediction: since $\lambda = 0.4$ lies in the regime where almost perfect pair transmission is predicted.

The bottom-right panel in Fig.~\ref{fig:timecutx} shows the dynamics for $\lambda = 6$, where Eq. \ref{Eq:pairexacttrans} would predict a strong suppression of the transmission. For these intermediate values of the frequency \( \hbar\omega \approx 3.82 J \), we also observe a marked suppression of pair transmission in our numerical results. At intermediate times ($tJ/\hbar \leq 60$), the wave packet is perturbed strongly at the impurity, showing large distortions around the impurity region. At later times ($tJ/\hbar = 120$), the wave packet has been reflected and only a very low transmission has taken place.
The deviations between the total density and the pair density are similarly low as in the case of no driving, which means that the impurity during the reflection does not break many of the pairs. 

The most intriguing behavior arises at $\lambda = 2.6$, shown in top-right of the Fig.~\ref{fig:timecutx}. At this values of driving, we see a partial transmission and enhanced dissociation. As the wave packet approaches the impurity (e.g., $tJ/\hbar = 60$), a noticeable mismatch begins to emerge between the total and pair densities. By later times ($tJ/\hbar= 120$), this difference becomes more pronounced and reflects the breakup of bound pairs. The observed differences between single and paired densities arises near the impurity. Once a pair is broken, there seem to occur two different processes. The first process is that part of the resulting single-particle component propagates with a higher group velocity \( v_s \propto J \), while the pair moves much more slowly \( v_p \propto J^2/|U| \). As seen in Fig. ~\ref{fig:U-10-density} 
this difference in velocities leads to the spatial separation as shown in the time evolution of the total density $\langle n_j(t) \rangle$, the pair density $\langle n_j(t)(n_j(t) - 1) \rangle$, and their difference
\[
n_{\text{single}}(j, t) = \langle n_j(t) \rangle - \langle n_j(t)(n_j(t) - 1) \rangle.
\]

Interestingly the second process observed is that additionally to the two propagating excitations, there is a long lived difference arising exactly at the impurity (see Fig. ~\ref{fig:U-10-density} (c)). This difference corresponds to singles that seem to remain dynamically trapped up to $tJ/\hbar\approx 200$. Meanwhile, the rest of the density continues to propagate, representing the untrapped part of the initial pair. We interpret this effect as particles entering the impurity as pairs, dissociating at the impurity, and then being trapped, since the single particle hopping between the impurity and the remainder of the system is suppressed. This (quasi-) localization of the single-particles, despite the presence of attractive interactions and available continuum states, is reminiscent of a \emph{Floquet bound state in the continuum} (BIC)~\cite{LonghiDellavalle2013,DellavalleLonghi2014}. While a full spectral analysis is beyond the scope of this work, the observed long-lived localization suggests the emergence of BIC-like behavior under periodic driving.

\section{Summary}
\label{sec:Conclusion}

In this paper we investigate transport in a one-dimensional Bose-Hubbard chain in the presence of a time-periodically modulated impurity. We systematically analyze the dynamics of both single particles and pair (doublons) injected into an empty lattice. A key focus lies in exploring the impact of the impurity's driving amplitude and frequency. Our findings confirm that, in the high-frequency regime, the transmission of pairs aligns well with Floquet-Schrieffer-Wolff predictions, including suppression of the transmission at certain driving strength. At intermediate and low frequencies we find that the results deviate considerably from the Floquet-Schrieffer-Wolff predictions and numerical analysis is needed. 	The density plots and pair correlation functions reveal rich dynamics, including wave packet spreading, trapping, and dissociation, with strong dependence on the interaction strength $U$ and driving parameters. At intermediate modulation strengths (e.g., $\lambda = 2.6$), pair breaking occurs: a part of the wave can remain trapped near the impurity, while the other part is reflected or transmitted. We find that a long localization of the single-particle at the impurity site can occur which  resembles findings for Floquet bound state in the continuum. This study bridges theoretical predictions and numerical simulations, highlighting bosonic analogs revealing new regimes of transport control via impurity driving. The results are experimentally relevant for ultracold atoms in optical lattices, where controlled impurity modulation can probe transport, pairing, and non-equilibrium dynamics in strongly correlated systems.

\emph{Acknowledgments:}
We thank F. H\"ubner, I. Schneider, and S. Eggert for fruitful discussions.
We acknowledge support by the Deutsche Forschungsgemeinschaft (DFG, German Research Foundation) under Project No.~277625399-TRR 185 OSCAR (``Open System Control of Atomic and Photonic Matter'', B3), No.~277146847-CRC 1238 (``Control and dynamics of quantum materials'', C05), CRC 1639 NuMeriQS (``Numerical Methods for Dynamics and Structure Formation in Quantum Systems'') – project
No.~511713970 , 
and under Germany’s Excellence Strategy – Cluster of Excellence Matter and Light for Quantum Computing (ML4Q) EXC 2004/1 – 390534769. 

\emph{Data availability:}
The supporting data for this article are openly available at Zenodo \cite{AhmadiSheikhan2025}.

\end{document}